\documentclass[%
 reprint,
 amsmath,amssymb,
 aps, pre, showkeys
]{revtex4-2}

\usepackage{graphicx}
\usepackage{dcolumn}
\usepackage{bm}
\usepackage[hidelinks]{hyperref}

\usepackage[utf8]{inputenc}
\usepackage[T1]{fontenc}
\usepackage[english]{babel}
\usepackage[varg]{newtxmath} 
\usepackage{pgfplots, tikz}
\usepackage{color, accents, booktabs}
\pgfplotsset{compat=1.18} 
\begin{document}

\preprint{APS/123-QED}

\title{Gyrokinetic equilibria of high temperature superconducting magnetic mirrors}

\author{M. H. Rosen}
 \email{mhrosen@pppl.gov}
\affiliation{Department of Astrophysical Sciences, Princeton University, Princeton, NJ 08540, USA}
\affiliation{Princeton Plasma Physics Laboratory, Princeton, NJ 08540, USA}

\author{M. Francisquez}
\affiliation{Princeton Plasma Physics Laboratory, Princeton, NJ 08540, USA}

\author{J. Juno}
\affiliation{Princeton Plasma Physics Laboratory, Princeton, NJ 08540, USA}

\author{B. Terry-Mendoza}
\affiliation{Princeton Plasma Physics Laboratory, Princeton, NJ 08540, USA}
\affiliation{University of Puerto Rico, Mayagüez Campus, Mayagüez, PR 00681, USA}

\author{A. Hakim}
\affiliation{Princeton Plasma Physics Laboratory, Princeton, NJ 08540, USA}

\author{G. W. Hammett}
\affiliation{Princeton Plasma Physics Laboratory, Princeton, NJ 08540, USA}

\date{\today}

\begin{abstract}
High-temperature superconducting (HTS) magnets and other advances have led to renewed interest in magnetic mirrors for fusion energy.
The non-Maxwellian nature of mirror plasmas necessitates kinetic modeling to predict, optimize and design mirrors.
Explicit gyrokinetic full-f codes can be used to study instabilities and turbulent transport in tokamaks and mirrors, but they have been prohibitively expensive to integrate directly over the very long time scales required to compute kinetic plasma equilibrium.
We demonstrate that these studies are now feasible thanks to novel multiscale methods delivering a 25,000X speed-up. 
The resulting kinetic equilibrium, electrostatic potential, and ion confinement time are consistent with analytic theory. 
This transformative capability opens the door to a new way of obtaining equilibria for mirrors, and we discuss how this technique may also accelerate calculations for tokamaks and stellarators. 
The models presented in this article address critical multiscale problems in modeling magnetic mirrors, opening a new research avenue for equilibrium studies using an explicit continuum gyrokinetic code.
\end{abstract}

\maketitle


\section{Introduction}
\label{sec: introduction}

High temperature superconductors (HTS) have enabled higher magnetic field strengths for magnetic fusion energy (MFE) applications, renewing interest in axisymmetric mirror configurations \citep{Post1995, Simonen_Cohen_Correll_Fowler_Post_Berk_Horton_Hooper_Fisch_Hassam_et_al, Fowler_1981, Endrizzi_Anderson_Brown_Egedal_Geiger_Harvey_Ialovega_Kirch_Peterson_Petrov_etal._2023}. Predictive modeling of these systems is required for design and optimization. The distribution of particles in mirrors are inherently non-Maxwellian, making full-f kinetic models necessary \citep{Francisquez_Rosen_Mandell_Hakim_Forest_Hammett_2023, dorf2025semi, Tran_Frank_Le_Stanier_Wetherton_Egedal_Endrizzi_Harvey_Petrov_Qian_et, Tyushev_Smolyakov_Sabo_Groenewald_Necas_Yushmanov_2025, Caneses-Marin_Harvey_Petrov_Forest_Anderson_2025, Frank_Viola_Petrov_Anderson_Bindl_Biswas_Caneses_Endrizzi_Furlong_Harvey_et}. 
 Gyrokinetic (GK) models have been one of the workhorses of MFE transport and turbulence modeling, yet computing mirror equilibria (which is needed to later compute transport) was prohibitively expensive due to the time scale separation between parallel dynamics and collisional scattering \citep{Francisquez_Rosen_Mandell_Hakim_Forest_Hammett_2023}.

Historically, the time scale separation between parallel advection and collisional scattering motivated bounce averaging, which removes fast advection and yields tractable reduced models \citep{ bendaniel1962scattering, marx1970effects, Pastukhov_1974,  Najmabadi_Conn_Cohen_1984, Hammett:1986, Mauel:1984, harvey1992cql3d, Harvey_Petrov_Forest_2016, li2025approximating, Rosen_Sengupta_Ochs_Parra_Hammett_2025, Ochs_2025}. Bounce-averaged studies, such as \citet{Pastukhov_1974}, show that collisions play an essential role in mirror confinement by scattering ions and electrons into the loss cone, a process balanced by the electric field. Thus, initial-value solvers must integrate to collisional time scales to reach equilibrium. While bounce-averaged models are valuable, they make an assumption that the distribution function is zero for passing particles. In a symmetric simple mirror, the trapped/passing boundary, or loss cone, is located at
\begin{equation}
    \mu_{\rm LC} (\mathbf x,v_\parallel)  = \frac{\frac{1}{2}m_s v_\parallel^2 + q_s\left(\phi(\mathbf x) - \phi_m\right)}{B_{m} - B(\mathbf x)}, \label{eq: mirror loss cone}
\end{equation}
where $B_{m}$ is the maximum amplitude of the magnetic field $\mathbf{B}$, $\phi_{m}$ is the electrostatic potential at the location of $B_{m}$ (the mirror throat), $\mu = m_s v_\perp^2/2B$ is the magnetic moment, $m_s$ is mass, $q_s$ is charge, $v_\parallel = \mathbf v \cdot \mathbf B$, $\mathbf v = (v_\parallel, \mu)$ are velocity space coordinates, and $\mathbf x$ is guiding center position.
Passing particles are important as they set the dynamics of expanders -- regions of flaring magnetic field outside the coils where field lines intersect end-plates. Expanders are regions where critical plasma-material interaction (PMI) occurs, and whose electrical conductivity influences the effectiveness of potential- or vortex-biasing schemes used to stabilize harmful interchange instabilities \citep{Ryutov_Berk_Cohen_Molvik_Simonen_2011}. However, going beyond bounce averaging, treating both passing and trapped particles, requires confronting the scale separation between parallel advection and collisional scattering.

Bridging these timescales remains a major computational barrier. One common method used for kinetic simulations of mirror plasmas is particle-in-cell (PIC) \citep{Tran_Frank_Le_Stanier_Wetherton_Egedal_Endrizzi_Harvey_Petrov_Qian_et, Tyushev_Smolyakov_Sabo_Groenewald_Necas_Yushmanov_2025, Caneses-Marin_Harvey_Petrov_Forest_Anderson_2025}.
However, PIC converges slowly (${\mathrm{error}} \propto 1/\sqrt{N}$, where $N$ is the total number of particles), and density decreases by a factor of $10^{6}$ from the center of the mirror (midplane) to the end-plate wall in the configuration studied here. 
To maintain a minimum count of particles per cell, large numbers of particles are needed for PIC. 
Continuum methods converge more rapidly than PIC and have fewer issues treating problems with large ranges in density; however, they require a mesh in velocity space, which often includes unimportant regions of phase space (large accelerations near $B_{m}$ and low $f$)\citep{Francisquez_Rosen_Mandell_Hakim_Forest_Hammett_2023, dorf2025semi}, imposing tiny time steps in explicit time integration.

Explicit methods are the simplest for advancing the solution of initial value problems in time, with their time step set by the Courant-Friedrichs-Lewy (CFL) condition.
Implicit methods have long been successful for diffusion-dominated problems, but can be complicated to implement and, until recently, had difficulty achieving net speedup for problems dominated by hyperbolic/advection dynamics. However, recent work using advances in multigrid methods \citet{Manteuffel:2018, Manteuffel:2019} has demonstrated an implicit method for the gyrokinetic mirror problem that achieves very large speedup \citep{dorf2025semi}.
Here, we present results from a novel multiscale pseudo orbit-averaging (POA) algorithm that achieves even larger speedup factors for the gyrokinetic mirror problem. This algorithm, first presented by us in \citet{rosen2025APseudoOrbitAveraging}, is relatively easy to implement in explicit time-stepping codes.
In addition, phase-space time-dilation methods can relax the CFL condition in unimportant regions of phase space \citep{Werner_Jenkins_Chap_Cary_2018, hopkins2025time}. 

This article leverages the computational gains from multiscale explicit methods to predict the equilibrium of HTS axisymmetric mirror configurations with continuum GK, demonstrating first-of-a-kind results. Section \ref{sec: model} describes the computational framework and simulation parameters. Section \ref{sec: results} presents results demonstrating that the simulation has reached steady state. Results are analyzed in section \ref{sec: discussion} to show agreement with analytic theory.
\section{Model description} \label{sec: model}

This study solves a GK equation of the form \citep{Idomura_Ida_Kano_Aiba_Tokuda_2008}
\begin{equation}
\frac{\partial f_s(\mathbf x , \mathbf v)}{\partial t} = \{\mathcal{H},f_s\} + C(f_s) + S, \label{eq: Full vlasov equation}
\end{equation}
where $f_s(\mathbf x , \mathbf v)$ is the guiding center distribution function of species $s$, and $t$ is time. The Poisson bracket $\{\mathcal{H},f_s\}$ of the Hamiltonian $\mathcal{H}$
\begin{equation}
    \mathcal{H} = \frac{1}{2}m_s v_\parallel ^2 + \mu B + q_s \phi
\end{equation}
with $f_s$ models particle transit parallel to $\mathbf{B}$ as well as acceleration by electrostatic and mirror forces. Ion-ion and ion-electron collisions are modeled in $C(f_s)$ using a Dougherty operator with collision frequencies dependent on local instantaneous parameters~\citep{Francisquez_Bernard_Mandell_Hammett_Hakim_2020, Francisquez_Juno_Hakim_Hammett_Ernst_2022}, and $S$ models sources. Equation~\ref{eq: Full vlasov equation} is solved for deuterium ions using a Boltzmann electron response \citep{Francisquez_Rosen_Mandell_Hakim_Forest_Hammett_2023}. The Boltzmann electron approximation is reasonable between the coils, where the ambipolar potential builds up and subsequently determines the ion confinement time \citep{Post1995}.

These simulations build upon the work of \citet{Francisquez_Rosen_Mandell_Hakim_Forest_Hammett_2023, Francisquez_Cagas_Shukla_Juno_Hammett_2025}, modeling the Wisconsin High-Field Axisymmetric Mirror (WHAM) experiment, which utilizes HTS magnets to reach fields with $B_{m} = 17 \; \rm T$ \cite{Endrizzi_Anderson_Brown_Egedal_Geiger_Harvey_Ialovega_Kirch_Peterson_Petrov_etal._2023}. We focus on the field line passing through $(r,z)=(0.1 \; \mathrm{m}, 0)$, reaching a minimum field at the midplane ($z=0$) of $B_0 = 0.53$ T, entailing a mirror ratio of $R_m = B_m/B_0 = 32$ \citep{Francisquez_Rosen_Mandell_Hakim_Forest_Hammett_2023, dorf2025semi}. The experiment's target parameters include hot deuterium ions ($T_i = 10$ keV) and electrons ($T_e = 940$ eV). Their target density is around $n_0 = 3\times 10^{19} \; \rm m^{-3}$. The ion thermal speed $v_{\rm ti} = \sqrt{T_i/m_i}$ is based on the temperature $8.36 \; \rm keV$, such that the parallel transit time is $\tau_{\parallel} =L/(2 v_{\rm ti}) = 3.17 \times 10^{-6}$ s ($L=2 \; \rm m$ is the distance between coils) while the ion-ion collision time at the midplane is $1/\nu_{ii} = 0.071$ s, a separation of four orders of magnitude. Making matters worse, the extreme mirror forces produced by HTS coils limit the timestep to $\Delta t \approx 5\times 10^{-11}$ s -- numerical integration must span nine orders of magnitude in time to reach steady state.

This extreme temporal scale separation is reconciled by a novel, explicit time integration method that is inspired by bounce-averaging techniques \citep{rosen2025APseudoOrbitAveraging}. The pseudo orbit-averaging (POA) scheme alternates between two phases. A full dynamics phase (FDP) solves equation \eqref{eq: Full vlasov equation} for a few transit times, allowing rapid equilibration of passing particles. An obit-averaged phase (OAP) separates the distribution function into regions of passing and trapped particles, freezes passing particles and slows down advection for trapped particles, allowing large time steps and equilibration between sources and collisions. The OAP solves
\begin{equation}
\frac{\partial f_s}{\partial t} = H_{\text{trap}}\left(\alpha\{\mathcal{H},f_s\} + C(f_s) + S\right),
\label{eq: dfdt orbit averaged phase}
\end{equation} 
where
\begin{equation}
H_{\text{trap}} = 
\begin{cases}
1, & \text{for trapped particles},\\[5pt]
0, & \text{for passing particles}.
\end{cases} \label{eq: trapped vs passing step function}
\end{equation}
and $\alpha <1$. The value of $\alpha$ is determined by requiring that $\alpha\{\mathcal{H},f_s\}$ and $C(f_s)$ impose the same CFL constraint on the time step. The step function $H_{\text{trap}}$ is computed based on the Yushmanov potential $\mu B + q_s\phi$, calculating trapped and passing regions numerically, evolving in time through its dependence on $\phi$. For this reason, our POA simulations perform multiple OAP-FDP cycles, in contrast to the reduced problems in \citet{rosen2025APseudoOrbitAveraging}, whose exact and time-independent trapped/passing boundary made optimal results with a single OAP-FDP pair possible. 

In addition, the time dilation schemes detailed in \citet{Werner_Jenkins_Chap_Cary_2018} and \citet{hopkins2025time} are employed, although applied to the novel setting of continuum GK in mirrors. Equation \eqref{eq: Full vlasov equation} changes to 
\begin{equation}
\frac{\partial f_s}{\partial t} = \beta(\mathbf x, \mathbf v) \left[\{\mathcal{H},f_s\} + C(f_s) + S \right], \label{eq: time dilated vlasov equation}
\end{equation}
where $\beta(\mathbf x, \mathbf v) \in (0,1]$ is a phase-space dependent factor effectively dilating time. The factor $\beta(\mathbf x, \mathbf v)$ is chosen to dilate time in the FDP by the amount necessary to increase the time-step by $10$, a magnitude chosen to locate unimportant regions of phase space. That is,
\begin{equation}
    \beta (\mathbf x, \mathbf v)= \min \left(1, \frac{\max(\omega_{\rm CFL}(\mathbf x, \mathbf v))/10}{\omega_{\rm CFL}(\mathbf x, \mathbf v)} \right),  \label{eq: time dilation factor}
\end{equation}
where $\omega_{\rm CFL}(\mathbf x, \mathbf v)$ is a local estimate of the magnitude of the dominant eigenvalue of the $\partial_tf_s$ operator in equation~\ref{eq: Full vlasov equation}. In these simulations, the maximum $\omega_{\rm CFL}$ is due to large gradients in $B$ near the mirror throats and small cells in $v_\parallel$ grids, so $\beta$ is independent of time \citep{Francisquez_Rosen_Mandell_Hakim_Forest_Hammett_2023}. A simulation without the time dilation scheme shows negligible differences in a POA study.
Equation~\ref{eq: time dilated vlasov equation} does not share the conservation properties of equation~\ref{eq: Full vlasov equation}, but the time dilation factor in equation~\ref{eq: time dilation factor} keeps the conservation errors below acceptable levels and achieves a considerable speedup.
\citet{Francisquez_Rosen_Mandell_Hakim_Forest_Hammett_2023} presented a similar conservative framework for increasing the timestep using a modified Hamiltonian; but modifying $\mathcal H$ modifies the equilibrium, while equation~\eqref{eq: time dilated vlasov equation} has the same equilibrium as equation~\ref{eq: Full vlasov equation}. Future studies could pursue using a $\beta(\mathcal{H})$, i.e. a function of Hamiltonian, as a means to preserve the conservative properties of the original system. Additionally, the FDP uses the low-pass filter described in \citet{rosen2025APseudoOrbitAveraging}, with a timescale of $5\times 10^{-6}\,\rm s$.

Initial conditions are constructed to be a Maxwellian distribution with density and temperature $(n, T) = (10^{17} \, \rm{m}^{-3},\, 830 \, \rm{eV})$ and parallel mean speed
\begin{align}
    u_{\rm \parallel} &= 
        \begin{cases}
            0, & \text{if } |z| < z_m \\
            7\frac{|z|}{z}\sqrt{\frac{T_e}{m_i}}\tanh\left( 6.06 \left|z - z_m\right| \right), & \text{otherwise.}
        \end{cases} 
\end{align}
Two alternative sources ($S$) are used -- one is Maxwellian inspired by \citet{Francisquez_Rosen_Mandell_Hakim_Forest_Hammett_2023}, and the other is a neutral beam injection (NBI) source inspired by \citet{dorf2025semi}. The source parameters are defined to produce WHAM-like results that match the target parameters of the experiment \citep{Endrizzi_Anderson_Brown_Egedal_Geiger_Harvey_Ialovega_Kirch_Peterson_Petrov_etal._2023}. A $\mathcal{E}_{\rm NBI} = 25 \, \rm keV$ beam with 200 eV spread is used, commensurate with prior studies \citep{dorf2025semi, Endrizzi_Anderson_Brown_Egedal_Geiger_Harvey_Ialovega_Kirch_Peterson_Petrov_etal._2023}. The total particle flux is $\int J S ~\mathrm{d}\mathbf{v}~\mathrm{d}z = 3.51\times10^{20} \rm m^{-3}s^{-1}$.

These sources must be distributed (i.e. bounce-averaged) along trapped particle orbits to avoid errors in the steady state reached by the OAP~\cite{rosen2025APseudoOrbitAveraging}. Therefore, the beam source is defined at the midplane and mapped along $z$ via energy conservation. At the midplane, the beam source $S_b$ is \citep{dorf2025semi}
\begin{align}
    S_{b,0}(v_{\parallel,0},\mu) = \Gamma_0 \exp \left(- \frac{\left( |v_{\parallel,0}| - v_{b} \right)^2 + \left( \sqrt{\frac{2 \mu B_0}{m_i}} - v_b \right)^2}{2 T_{b}/m_i} \right)
\end{align}
which is mapped along $z$ as
\begin{equation}
    S(z, v_\parallel,\mu) = S_0\left(\sqrt{v_\parallel^2 + 2 \mu \left(B\left(z\right) - B_0 \right)/m_i}, \mu\right).
\end{equation}
Here, $\Gamma_0 = 200 \; \rm s^{-1}\, m^{-3}$, $T_b = 200 \; \rm eV$, $v_b = \sqrt{\mathcal{E}_{\rm NBI} / m_i}$. The Maxwellian source is constructed in a similar fashion, except that the source at the midplane is defined as Maxwellian.
\begin{align}
    S_{\rm max, 0}(v_{\parallel,0},\mu) =  \frac{\Gamma_{\rm max}}{2 \pi T_{\rm max}/m_i} \exp \left(- \frac{v_{\parallel,0}^2 + \frac{2 \mu B_0}{m_i}}{2 T_{\rm max}/m_i} \right)
\end{align}
where $(\Gamma_{\rm max}, T_{\rm max}) = (4.23\times10^{7} \rm s^{-2}\, m^{6},\,19.9 \, \rm keV)$.

The phase space grid is constructed using non-uniform velocity and position space mappings \citep{Francisquez_Cagas_Shukla_Juno_Hammett_2025}. 
The map in $\mu$ is linear below $\mu = \mu_{max}0.2^2$ and cubic up to $\mu = \mu_{max}$, where $\mu_{\max} = m_i(3 v_{ti})^2/2B_0$ is the maximum extent of the grid, which balances packing cells near the loss cone with not making cells too small. The map in $v_\parallel$ uses $\tan(b\bar v _\parallel) / \tan(b)$ for $b=1.4$ where $\bar v_\parallel \in (-1,1)$, which packs cells near $v_\parallel = 0$ with large cells at large $v_\parallel$. Grid extents are $\pm16 v_{ti}$ in $v_\parallel$.
The map in $z$ packs cells in regions of large $\partial_z B$. 
The grid uses (400, 64, 32) cells in $(z,v_\parallel,\mu)$ using a hybrid polynomial basis \cite{Francisquez_Cagas_Shukla_Juno_Hammett_2025}. Each OAP lasts $0.1 \; \rm  s$ ($1.41\;\nu_{ii}^{-1}$) with $\alpha = 2 \times 10^{-5}$ and each FDP lasts for $15 \; \rm \mu s$ ($4.73\; \tau_{\parallel}$) with the final FDP extended to $60 \; \rm \mu s$. OAPs are performed, with an FDP following, the toal amount of time spent in the pseudo-orbit averaging scheme is $t_{\rm POA} = 2.05\nu_{ii}^{-1} + 37.8 \tau_\parallel$.
\section{Results} \label{sec: results}

\begin{figure}
    \centering
    \includegraphics[width=\linewidth]{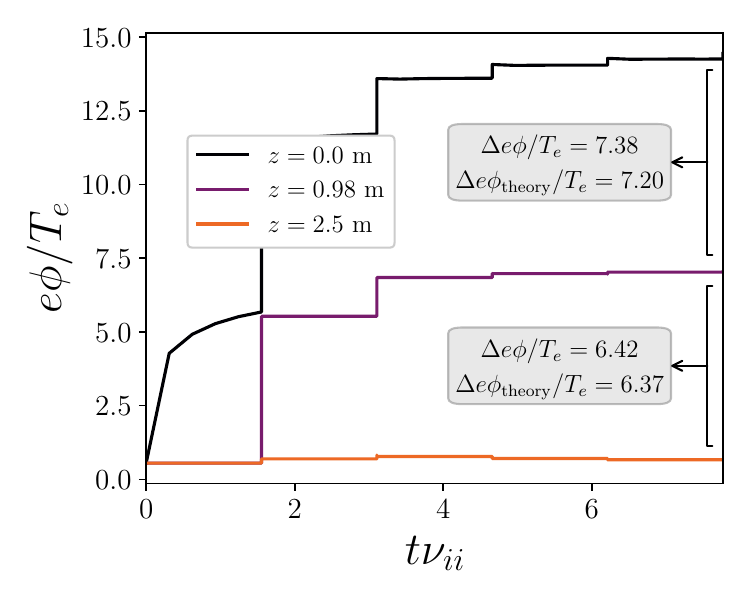}
    \caption{Time trace of the electrostatic potential of the Maxwellian source simulation at the midplane ($z=0$), mirror throat ($z=0.98$) and sheath entrance ($z=2.5$). The grey legend compares potential differences between the simulation and theory \citep{Rosen_Sengupta_Ochs_Parra_Hammett_2025}.}
    \label{fig: gkeyll field}
\end{figure}

\begin{figure*}
    \centering
    \includegraphics[width=\textwidth]{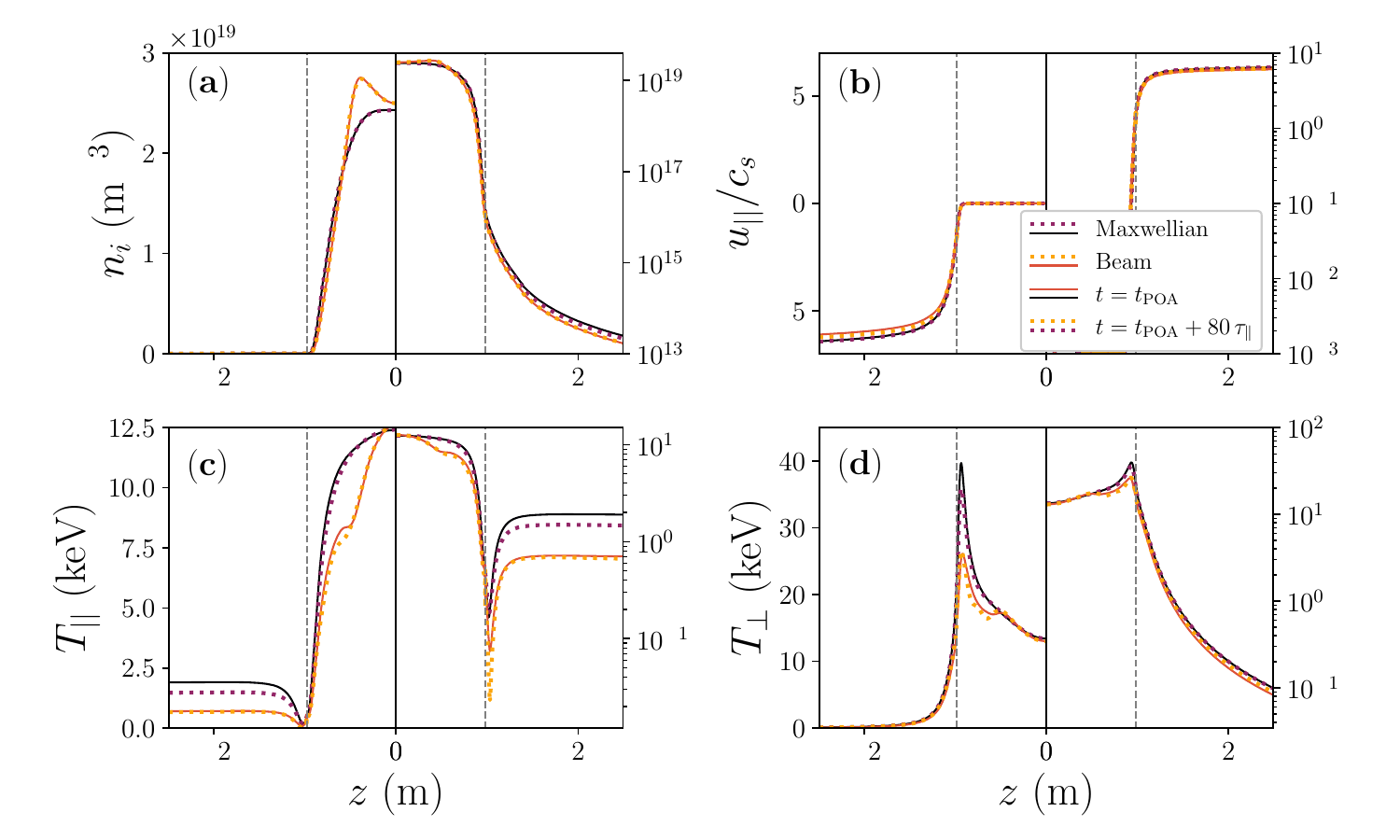}
    \caption{A comparison of the ion density (a), parallel bulk speed normalized to $c_s=\sqrt{T_e / m_i}$, parallel (c) and perpendicular (d) temperatures in the steady state of the Maxwellian (black and purple) and NBI (red and orange) sourced simulations. Results are symmetric about $z=0$, and we show the right half ($z>0$) in logarithmic scale. Grey dashed lines mark the mirror throats. The equilibrium produced by the simulation (solid lines) does not change much if evolved further using an FDP without time dilation (dotted lines). }
    \label{fig: gkeyll bimaxwellain}
\end{figure*}


The simulation with a Maxwellian source is presented in figure \ref{fig: gkeyll field}, which shows the time evolution of the electrostatic potential at the center ($z=0.0$ m), the mirror throat ($z=0.98$ m), and the sheath entrance ($z=2.5$ m). The plateau in the evolution of electrostatic potential at late times indicates that these simulations reach steady state after a few ion collision times, as expected. Brackets indicate that the drops in potential between the chosen points agree with theoretical estimates \citep{Rosen_Sengupta_Ochs_Parra_Hammett_2025}.

\begin{figure*}
    \centering
    \includegraphics[width=0.9\textwidth]{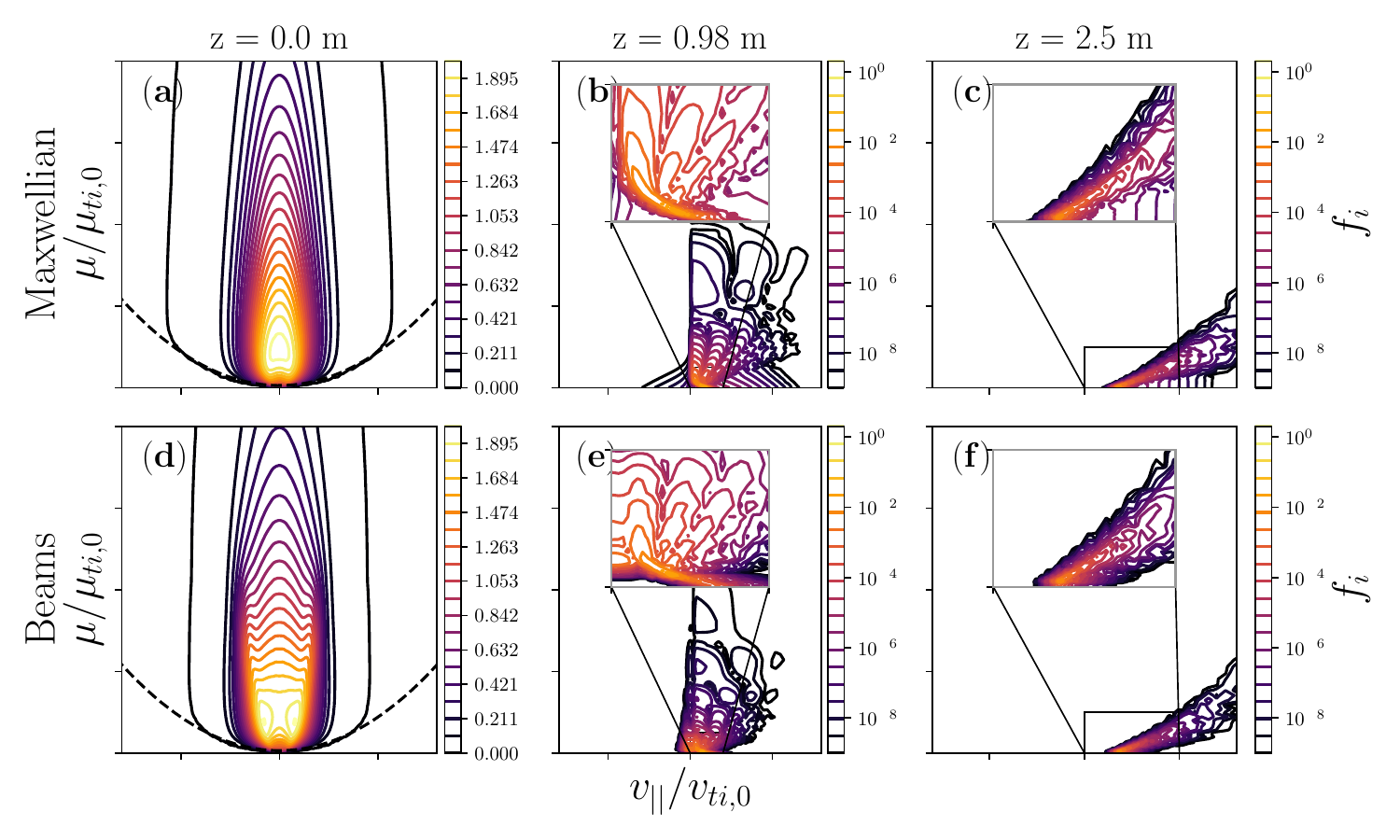}
    \caption{
    Equilibrium ion distribution function in the Maxwellian (top) and NBI (bottom) sourced simulations at the center (a,d), mirror throat (b,e) and sheath entrance (c,f); negative cells multiplied by -1 for ease of viewing. The white dashed line indicates the trapped-passing boundary, defined in equation \eqref{eq: mirror loss cone}.}
    \label{fig: gkeyll distf}
\end{figure*}

Moments of the distribution function (density $n$, parallel bulk speed $u_\parallel$, parallel temperature $T_\parallel$, and perpendicular temperature $T_\perp$) at steady state are shown in figure \ref{fig: gkeyll bimaxwellain} ($t = t_{\rm POA}$). In addition, an FDP-only continuation of the simulation ($t=t_{\rm POA} + 80\tau_\parallel$), is shown to demonstrate that the results of the POA simulation do not change when solving equation \eqref{eq: Full vlasov equation} directly, demonstrating that it is a solution near equilibrium. The moments are commensurate with the results presented in \citet{Francisquez_Rosen_Mandell_Hakim_Forest_Hammett_2023}, although this simulation has run $5000 \times$ longer.

As expected, at equilibrium, both the Maxwellian and NBI-sourced plasmas are confined by magnetic forces, with a density peak between the coils and exponential decay in the expanders. The NBI-sourced simulation, however, exhibits enhanced confinement with larger density and off-center maxima due to the off-center turning points of high-energy ions \citep{Ryutov_Berk_Cohen_Molvik_Simonen_2011}. This leads to a small confining potential of $e\Delta\phi/T_e=0.0977$. Ions are accelerated as they leave the mirror, reaching speeds near the sound speed at the mirror throat, and continue accelerating until they enter the sheath. The parallel temperature is reduced near the mirror throat, with the perpendicular temperature peaking due to $\mu$ conservation. In the expander, the parallel temperature plateaus and the perpendicular temperature decays. The beam source produces lower temperatures in the confined region because the beam’s spread is narrower than the Maxwellian distribution. The qualitative features of the moments of the distribution function agree with those presented in \citet{Wetherton_Le_Egedal_Forest_Daughton_Stanier_Boldyrev_2021} and \citet{Tyushev_Smolyakov_Sabo_Groenewald_Necas_Yushmanov_2025}, which analyzed similar models for a mirror-trapped Maxwellian source. 

In order to better understand the moments in figure \ref{fig: gkeyll bimaxwellain}, the distribution function offers insight.
Figure \ref{fig: gkeyll distf} compares the distribution function between the Maxwellian and beam sources at key points in space. The loss boundary is depicted in white and is reproduced well in these figures by both sources. The beams display a bimodal distribution (figure \ref{fig: gkeyll bimaxwellain}c), consistent with prior work in \citet{dorf2025semi} and \citet{Frank_Viola_Petrov_Anderson_Bindl_Biswas_Caneses_Endrizzi_Furlong_Harvey_et}. At $z=0.98$, all particles transition to free-streaming with positive parallel velocity. They are accelerated into the sheath at $z=2.5 \, \rm m$.

The transformative speed-up provided by the POA method enables parameter scans and predictions of confinement quality, enabled by HTS technologies that permit higher $R_m$. To that end, we present a scan of $R_m$ (holding $B(z=0)$ constant) in figure \ref{fig: magnetic field scan and confinement time}, measuring ion particle confinement time $\tau_p$ for each configuration.
For efficiency, the lower $R_m$ simulations $R_m=3$ and $R_m=5$ use resolutions of $192$ and $256$ cells in $z$, respectively, while $R_m=50$ uses $800$ cells in $z$. The linear fits for both sources agree closely with the simulation results, yielding $R^2$ values of $0.9995$ for the Maxwellian source and $0.9976$ for the beam. A simulation using a kinetic electron model (\citep{Francisquez_Rosen_Mandell_Hakim_Forest_Hammett_2023}) is included to show that the ion confinement time does not change significantly, as the Boltzmann electron response is reasonable between the field coils.
\section{Discussion} \label{sec: discussion}

\begin{figure}
    \includegraphics[width=\linewidth]{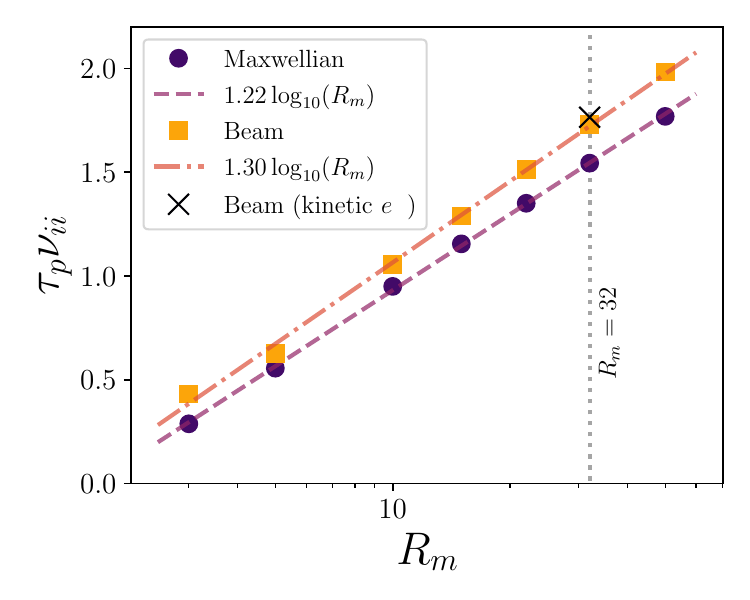}
    \caption{ Measuring the ion confinement time $\tau_p$, normalized to the ion-ion collision time $\nu_{ii}$ as a function of mirror ratio $R_m$.}
    \label{fig: magnetic field scan and confinement time}
\end{figure}

The calculation of GK equililbria in mirrors with a code based on direct explicit time integration is made possible by the transformative acceleration provided by the new POA algorithm. For the $R=32$ beam and Maxwellian simulations, to reach the same simulation time of $0.5$ seconds, a purely FDP simulation would need $18.9$ years, while the $R=32$ POA simulation took $6.7$ hours (on 8 NVIDIA A100 GPUs), leading to a speedup of \textbf{$\sim 25,000$}. Time dilation itself accounts for a $5\times$ speedup, yielding $5,000\times$ from POA. Lower mirror ratio simulations are much cheaper, with the $R=3$ case in figure \ref{fig: magnetic field scan and confinement time} taking 45 minutes on 8 NVIDIA A100 GPUs.

This performance is achieved alongside algorithmic choices tailored to accurately resolve the phase-space structure during the OAP. When determining $H_{\rm trap}$, the corners of each cell are located and labeled.
If any corner of a cell is labeled transiting, the entire cell is labeled as a transiting region of phase space.
During the OAP, a factor of $5\times 10^{-6}$ balances the collisional time-step with the advective time step. 
The system is in the region of distributed source/localized sink, as explored in \citet{rosen2025APseudoOrbitAveraging}, because the loss cone is closer to the beam at the midplane than at the turning points.
This scenario poses no issue; however, it requires a slightly higher alpha than optimal and more time per FDP, so the alpha is increased by a factor of 4 to $\alpha = 2 \times 10^{-5}$. 
A gap centered around the FDP’s steady state forms during the OAP and quickly decays during the FDP due to the low pass filter.
Additional speed-up and accuracy could be achieved through taking an exact orbit average of the distribution function during the OAP.


\subsection{Equilibrium potential compared to theory}
\label{sec: comparing the potential with theoretical predictions}

An important feature of any equilibria predicted for a mirror is the potential drop between the center and the throat, as this potential provides an important means to confine electrons and reduce rapid electron heat losses. Examining the final state of the Maxwellian source simulation, this potential drop is $7.38 \;\Delta e \phi/T_e$. The theoretical value for the potential equates the electron and ion confinement times. The ion particle confinement time is measured as
\begin{equation}
    \tau_{p} = \frac{\int_{-z_m}^{z_m} J f \; \mathrm{d}\mathbf{v}~\mathrm{d}z}{\int_{-z_m}^{z_m} J S\;  \mathrm{d}\mathbf{v}~\mathrm{d}z}. \label{eq: ion ion confinement time}
\end{equation}
The electron particle confinement time is estimated in \citet{Pastukhov_1974}, \citet{Najmabadi_Conn_Cohen_1984}, and \citet{Rosen_Sengupta_Ochs_Parra_Hammett_2025}. Using the confinement time from in \citet{Rosen_Sengupta_Ochs_Parra_Hammett_2025} ($c_0 = 1.117$, $Z_s = 1.0$), a value of $\Delta e \phi /T_e = 7.20$ is obtained, which is $2.55\%$ lower than the simulation. The results of both \citet{Pastukhov_1974} and \citet{Najmabadi_Conn_Cohen_1984} provide estimates of $\Delta e \phi /T_e = 6.04$ and $\Delta e \phi /T_e = 5.99$, respectively, which are further away from the value calculated in the simulation.
In this case, \citet{Rosen_Sengupta_Ochs_Parra_Hammett_2025} is the more relevant model because of its use of a Dougherty collision operator.

In the expander, the simulation indicates a drop of $6.42 \;\Delta e\phi/T_e$. The theoretical estimate is based on the adiabatic expansion of ions and conservation of magnetic flux \citep{Ryutov_2005, Skovorodin_2019, Gupta_Yushmanov_Barnes_Dettrick_Onofri_Tajima_Binderbauer_TAE_Team_2023}.
\begin{equation}
    \frac{\Delta e \phi}{T_e}  = \ln \left(\frac{B_{m} u_{\parallel}(z) }{ B(z) u_{\parallel, \rm  m}}\right). \label{eq: potential theoretic in expander}
\end{equation}
Inserting the parameters of our simulation into equation \eqref{eq: potential theoretic in expander} gives a potential drop of $6.37 \; \Delta e \phi / T_e$, which is $0.779\%$ lower than the simulation result. Matching theoretical predictions of the ambipolar potential was challenging in \citep{Francisquez_Rosen_Mandell_Hakim_Forest_Hammett_2023} because those simulations could not run long enough and because theory using a Dougherty collision operator was not available. We overcame both of these challenges through new theoretical estimates with Dougherty collisions \citep{Rosen_Sengupta_Ochs_Parra_Hammett_2025} and a method to accelerate numerical calculations of GK equilibria \citep{rosen2025APseudoOrbitAveraging}.

\subsection{Verification of steady state equilibrium}
\label{sec: validation study using purely fdp}

We verify that the result of the POA simulation is indeed the equilibrium. To that end, we perform an FDP-only simulation in figure \ref{fig: gkeyll bimaxwellain}. The initial condition of the FDP-only simulation is the final state of the POA algorithm at $t = t_{\rm POA}$. If the FDP-only simulation doesn't drastically alter the plasma profiles or produce instabilities, we will conclude that the POA result is indeed a steady state. This simulation runs for $240 \; \rm \mu s$, about $80$ transit times. There is a negligible change to the plasma moments in the confined region, indicating that the final solution is at equilibrium. The temperature in the region outside the mirror throat decreases in time; the difference is small, but noticeable. Furthermore, $\int J\dot f d{\bf v}dz / \int J S d{\bf v} dz$ measures the distribution function's evolution relative to the source, with small values indicating a solution near equilibrium. The end of the POA scheme has a value of $4.02\%$, and after $80 \tau_\parallel$, this quantity is $3.18\%$. These results demonstrate that the POA simulation yields a good equilibrium around which we can perform additional analysis in the future, such as studying turbulent interchange transport with 3D simulations.

\subsection{Confinement time scaling law}
\label{sec: determination of ion ion confinement time}
A key parameter in determining fusion power output is the ion confinement time $\tau_p$. Several works determine the ion confinement time in a magnetic mirror based on an analytic analysis of the Rosenbluth Fokker-Planck operator, which is to first-order \citep{bing1961end, Baldwin_1977, Fowler_1981, Egedal_Endrizzi_Forest_Fowler_2022}
\begin{equation}
    \tau_{p} = \nu_{ii}^{-1}\kappa \log_{10} R_m, \label{eq: analytic ion confinement time}
\end{equation}
where $\kappa \sim 1.0-1.75$~\citep{bing1961end}, depending on the form of the source (Maxwellian vs beam), beam injection angle, injection energy, and ion temperature, to name a few. Both slopes from the linear fits in figure \ref{fig: magnetic field scan and confinement time} from the Maxwellian ($\kappa = 1.22$) and beam ($\kappa = 1.30$) sources are within the range reported in the literature, using the collision frequency from \citet{Francisquez_Bernard_Mandell_Hammett_Hakim_2020} at the midplane. 
These results further corroborate the conclusion that beams have a longer confinement time than a Maxwellian source. 

Another expression for the confinement time is given in \citet{Baldwin_1977} and \citet{Fowler_1981} for $50\%$ deuterium, $50\%$ tritium beam injection at $90^\circ$ to the magnetic field. They give $n_{20} \tau_p = \mathcal{C} \mathcal{E}_{\rm NBI, keV}^{3/2} \log_{10}R_m$ with $\mathcal{C} \simeq2.4-2.8\times 10^{-4} \, \rm s$, where $n_{20} =n /10^{20} \rm \, m^{-3}$ and $\mathcal{E}_{\rm NBI, keV}$ is the beam energy in units keV \citep{Baldwin_1977, Egedal_Endrizzi_Forest_Fowler_2022}. Using the midplane density, performing linear regression on the beam results produces $\mathcal{C} = 1.58\times10^{-4} \, \mathrm{s}$. This result is $34\%$ lower than the range provided in \citet{Baldwin_1977} and \citet{Fowler_1981}, and is a result of using beams at $45^\circ$. \citet{Egedal_Endrizzi_Forest_Fowler_2022} shows that confinement time decreases with beam angle and the $\mathcal{C}$ value measured here agrees with their results at $45^\circ$.
\section{Conclusion}
\label{sec: conclusion}

This work overcomes a longstanding barrier that prevented explicit gyrokinetic codes from directly calculating gyrokinetic (GK) equilibria, due to the severe timescale separation between parallel dynamics and collisions, by using a novel multiscale algorithm that is easy to implement in explicit codes. Using a high-field (17 T, $R_m = 32$) axisymmetric mirror configuration, we present the first GK equilibrium solutions using this POA algorithm that include the expander region and kinetic ions and electrons simultaneously, at modest computational cost.

Explicit multiscale methods are leveraged to achieve a $25,000 \times$ speed-up while maintaining good accuracy of the resulting equilibria. The computed ambipolar potential matches theory in the center and expander of the mirror with $2.55\%$ and $0.779\%$ accuracy, respectively.
A parameter sweep determines the dependency of the ion confinement time on mirror ratio, which falls within the range of previous calculations, reproducing a known result that neutral beam injection (NBI) leads to better confinement than a Maxwellian source. Although results are shown in equilibrium, transients can be obtained by reducing the amount of time spent in the orbit-averaged phase.

Accurate predictions of experimental conditions may benefit from additional rigor and more representative conditions. 
Additional terms and features would somewhat increase the cost and modify the equilibrium, however would not introduce a new timescale; some examples are collisions with neutrals, a more accurate collision operator, finite Larmor radius (FLR) effects, centrifugal rotation, modeling plasma-material interactions, and additional heating terms (e.g., ECRH, HHFW). Although the GK model does not include the drift-cyclotron loss cone (DCLC) instability, one could include a reduced model \citep{Tran_Frank_Le_Stanier_Wetherton_Egedal_Endrizzi_Harvey_Petrov_Qian_et}. 
These features are left to future work. 
One of the advantages of our approach is that we directly calculate the expander region as well, while some previous work either used a lowest-order bounce-averaged approximation that neglects the expanders or would have too much particle noise because of the extremely low density there.  We also included ion-electron collisions, which have been neglected in some recent work but are important for ion equilibria \citep{Baldwin_1977}.

These results demonstrate the POA algorithm in first-of-a-kind results, showing dramatic acceleration in solving full-f GK equilibria. Burning plasma tokamak and stellarator reactors present a similar problem -- distribution functions can have a strong mirror-trapped/passing component whose saturation is reached on the long timescale of weak collisions, but in the presence of fast parallel transport, which imposes a small time step \cite{Shukla_Roeltgen_Kotschenreuther_Juno_Bernard_Hakim_Hammett_Hatch_Mahajan_Francisquez_2025}. Particle-in-cell methods may also benefit from the POA algorithm, as they conceptually solve the same equations by different means (tracking virtual particles whose statistics constitute the distribution $f_s$). The key is to figure out whether a particular particle (or point in phase space) is on a periodic or transit (loss) orbit.


\begin{acknowledgments}
The authors are grateful for conversations and discussions with Antoine Hoffmann and Cary Forest. Generative AI was not used for the text of this article or the code that produced these results. This work was supported by Princeton University and the U.S. Department of Energy under contract number DE-AC02-09CH11466. The United States Government retains a non-exclusive, paid-up, irrevocable, world-wide license to publish or reproduce the published form of this manuscript, or allow others to do so, for United States Government purposes. Part of this work was funded by the Department of Energy, Office of Fusion Energy Science FIRE Collaborative: Fusion Neutrons for Integrated Blanket Technology Development Through Advanced Testing and Design, as part of the work to understand HTS mirrors as a neutron source (Award DE-SC0026040).
\end{acknowledgments}

\section*{Data availability statement}
Simulations presented in this manuscript can be reproduced using the open-source code \texttt{Gkeyll}~\citep{gkeyllGkeyllDocumentation} and the input files found at \url{https://github.com/ammarhakim/gkyl-paper-inp/tree/master/2026_PRE_hts_mirror_equilibria}



\bibliography{tex-files/bibliography}

\end{document}